# Conditional Randomization Tests for Behavioral and Neural Time Series

**Kenneth D. Harris[1*], Kevin J. Miller[1,2]**
1. UCL Queen Square Institute of Neurology, London WC1N 3BG, UK
2. Google DeepMind, London, UK
*Correspondence: kenneth.harris@ucl.ac.uk

Randomization tests allow simple and unambiguous tests of null hypotheses, by comparing observed data to a null ensemble in which experimentally-controlled variables are randomly resampled. In behavioral and neuroscience experiments, however, the stimuli presented often depend on the subject's previous actions, so simple randomization tests are not possible. We describe how conditional randomization can be used to perform exact hypothesis tests in this situation, and illustrate it with two examples. We contrast conditional randomization with a related approach of tangent randomization, in which stimuli are resampled based only on events occurring in the past, which is not valid for all choices of test statistic. We discuss how to design experiments that allow conditional randomization tests to be used.

Randomized experiments are considered a "gold standard" for scientific research because they allow simple and unequivocal tests of null hypotheses and can demonstrate causality [1]. If some experimental variables were generated at random, we can test the null hypothesis that these randomized variables have no effect on other measured variables using a randomization test. To do so, we compute a test statistic measuring the association of the randomized and measured variables, and compare this test statistic to a null ensemble obtained by repeatedly regenerating the randomized variables and recomputing the test statistic. If the randomized variables have no effect on the measured variables, then the test statistic follows the same probability law as the null ensemble. So if the test statistic lies outside the $\alpha^{th}$ percentile of the null distribution, we infer a causal effect at significance $\alpha$.

In behavioral or neural sciences, we often record a subject's responses to a random sequence of stimuli. To apply a randomization test for a causal effect of the stimuli on the responses, we must be able to randomly regenerate the stimulus sequence. If the stimulus presented on each trial is independent of the subject's previous responses, there is no problem. But if the stimuli depend on the subject's previous responses, we cannot randomly regenerate the entire stimulus sequence without also taking into account these previous responses.

Here we describe how conditional randomization tests can be used in this situation. Conditional randomization tests are simple, and have a history going back at least as far as RA Fisher. However, they have seen surprisingly little application until the last few years (see Refs. [3,4] and citations therein). Conditional randomization requires stimuli to be resampled conditional on the full sequence of past and future choices. We distinguish this from tangent randomization, in which each stimulus is resampled based only on past responses. Conditional randomization tests are exact: the probability of falsely rejecting the null hypothesis equals the significance level, even for small dataset sizes. Tangent randomization tests are not generally exact, but are sometimes approximately valid for large dataset sizes. We give two examples to show how conditional randomization tests can be used for behavioral timeseries, and how tangent randomization can sometimes fail.

### The conditional randomization test in general

Conditional randomization tests work when our null hypothesis provides a way to randomly resample some variables (such as stimuli) conditional on other variables (such as choices and rewards). Whether conditional randomization can be used depends on the exact experimental design and must be checked on a case-by-case basis.

Formally, let $\boldsymbol{X}$ be a vector summarizing all data in an experiment. For example, $\boldsymbol{X}$ could contain the full sequences of stimuli, choices, and rewards. In general, we will not have a null hypothesis for the full probability distribution $\mathbb{P}[\boldsymbol{X}]$ since we do not know the probability distribution governing the subject's responses. Nevertheless, we may be able to state our null hypothesis in the form of a conditional distribution $\mathbb{P}[\boldsymbol{X}|\boldsymbol{S}(\boldsymbol{X})]$, where the vector $\boldsymbol{S}(\boldsymbol{X})$ is some deterministic function or subset of the vector $\boldsymbol{X}$. For example, $\boldsymbol{S}(\boldsymbol{X})$ could contain the sequences of responses and rewards, but not stimuli. The conditional distribution $\mathbb{P}[\boldsymbol{X}|\boldsymbol{S}(\boldsymbol{X})]$ gives a recipe for resampling a sequence of stimuli consistent with past and future responses and rewards, which is valid when the null hypothesis is true.

To perform the test, we define a test statistic $T(\boldsymbol{X})$, for example measuring the correlation between the sequences of stimuli and responses. We create a null ensemble by repeatedly sampling values $\mathbf{X}'$ from the distribution $\mathbb{P}[\boldsymbol{X}|\mathbf{S}(\mathbf{X})]$, and define the p-value as the percentile of the actual test statistic $T(\boldsymbol{X})$ within the null ensemble $\{T(\boldsymbol{X}')\}$: if $T(\boldsymbol{X})$ is larger than $M$ of the $N$ random resamples, then $p = \frac{1+N-M}{1+N}$. (Ties are broken randomly, which can be achieved by adding a small random number to $T(\boldsymbol{X})$ and any test statistics equal to it.) If the null hypothesis is true, $\boldsymbol{X}$ is an unexceptional random sample from $\mathbb{P}[\boldsymbol{X}|\mathbf{S}(\mathbf{X})]$, so $\mathbb{P}[p \leq \alpha] = \alpha$, as required for an exact test.

One can use a conditional randomization whenever it is possible to find a conditioning statistic $\boldsymbol{S}(\boldsymbol{X})$ and an algorithm to sample from $\mathbb{P}[\mathbf{X}|\boldsymbol{S}(\boldsymbol{X})]$ under the null hypothesis. We now present two examples where this is the case.

### Example 1: reaction time task

Our first example is a reaction time task. The task a simplified version of one used in Ref. [5], where conditional randomization was used to determine the day on which mice learned to respond to a visual stimulus.

In the reaction time task, a subject responds to a sensory stimulus by pressing a button, after which a reward is delivered. To discourage a strategy of constant pressing, the stimulus only appears once the subject has refrained from pressing the button for a "quiescence interval" whose length is drawn randomly on each trial. A subject who can see the stimulus will do best by waiting for the stimulus to appear and then immediately responding. However, a subject who can detect rewards but not stimuli can still perform reasonably well by guessing: waiting after each reward for a time of approximately the mean quiescent period, then pressing the button.

Our null hypothesis is that the subject can detect rewards but not stimuli. We cannot use a simple randomization test, because this tests a stronger null hypothesis that the entire stimulus sequence is independent of the entire response sequence, which is false as stimulus times depend on previous responses. We can however use conditional randomization. To do so we must find a way to randomly resample the sequence of quiescence intervals conditional on the sequences of responses and rewards, in a manner which is only valid under the null hypothesis that the subject can detect rewards but not stimuli.

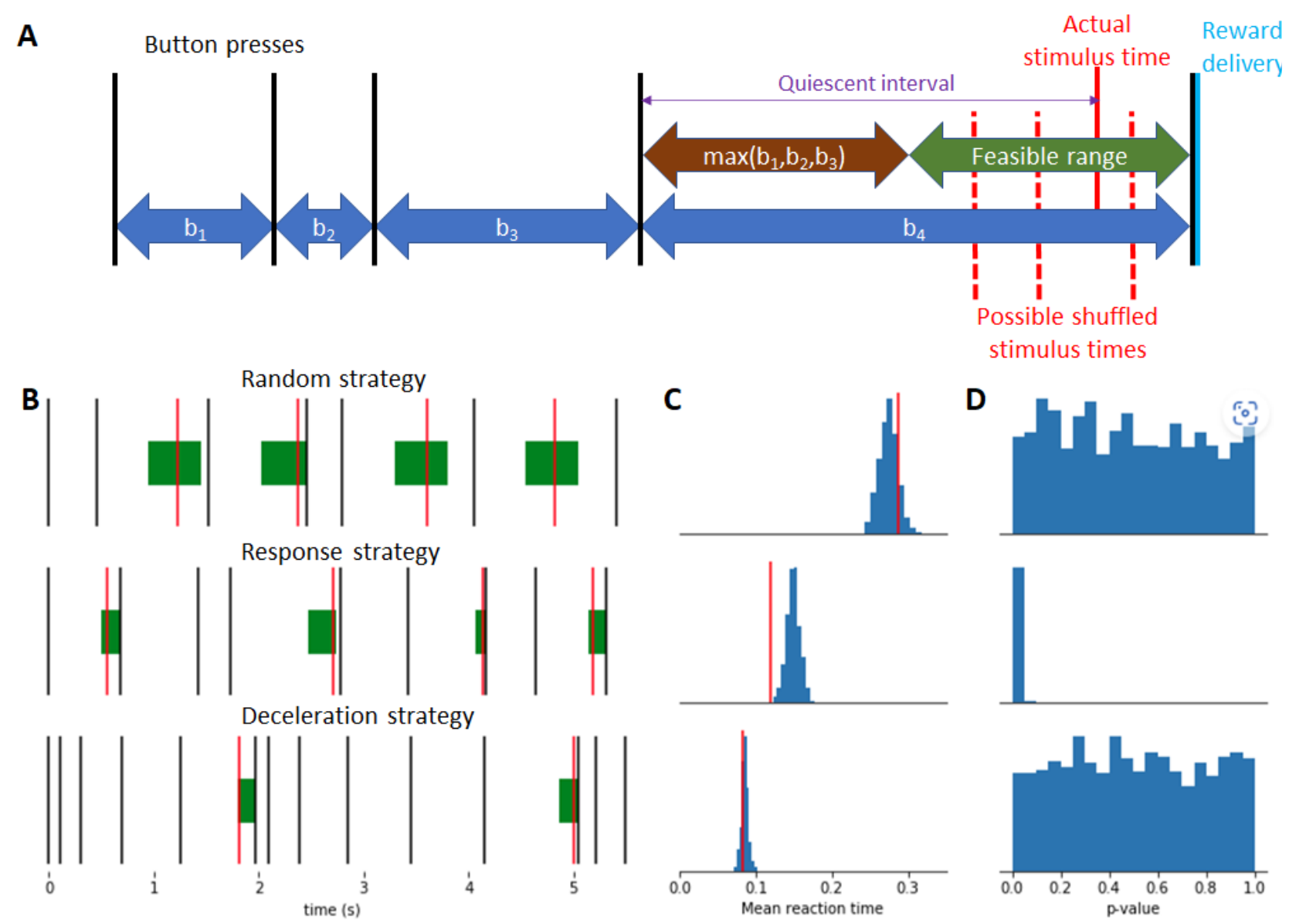

**Figure 1 | Conditional randomization analysis for reaction time task. A.** An example trial of the task. A stimulus appears after the subject has refrained from pressing the button for quiescent interval (purple), which is drawn independently on each trial from a uniform distribution between 500 ms and 1 s. The first three button press intervals $b_1 \ldots b_3$ were shorter than the trial's quiescent interval so a stimulus did not appear. Because interval $b_4$ was longer than the quiescence time, a stimulus appeared, a reward was delivered when the button was pressed, then a new trial began. To apply a conditional randomization test we randomly resample each trial's quiescence interval, truncated to a "feasible range" that would result in the same press being rewarded. The endpoints of this feasible range are the final inter-press interval and the longest previous inter-press interval. **B.** First few seconds of example simulations of subjects employing three possible strategies: pressing the button at random; responding to the stimuli with a variable delay; and a "deceleration strategy" of pressing at progressively longer intervals until a reward is delivered. Black lines: button presses; red lines: stimuli; green bars: ranges from which stimulus times are resampled, which are the intersection of the feasible range and the original 500 ms – 1s range. **C.** Red lines: mean reaction times for the three strategies. Blue histograms: null distribution of mean reaction time under conditional randomization. **D.** Histograms of p-values for 1000 simulations of each strategy. Significance is reliably observed for the response strategy but not for the strategies that do not use the stimulus times.

Formally, let $q_t$ be the length of the quiescent period on trial $t$, and let $\boldsymbol{Q}$ be a vector containing the sequence of $q_t$. Let $\boldsymbol{B}$ be a vector containing the sequence of button press times. The vectors $\boldsymbol{B}$ and $\boldsymbol{Q}$ have lengths the number of button presses and the number of trials, which will not be equal as there can be several presses per trial. Let $\boldsymbol{R}$ be a binary vector of the same length as $\boldsymbol{B}$, with an entry 1 if that button press followed a stimulus appearance and thus preceded a reward. Let $\boldsymbol{X} = \{\boldsymbol{Q}, \boldsymbol{B}, \boldsymbol{R}\}$ and $\boldsymbol{S}(\boldsymbol{X}) = \{\boldsymbol{B}, \boldsymbol{R}\}$, so sampling from $\mathbb{P}[\boldsymbol{X}|\boldsymbol{S}(\boldsymbol{X})]$ is equivalent to sampling from $\mathbb{P}[\boldsymbol{Q}|\boldsymbol{B}, \boldsymbol{R}]$.

By Bayes' theorem,

$$\mathbb{P}[\boldsymbol{Q}|\boldsymbol{B}, \boldsymbol{R}] = \frac{\mathbb{P}[\boldsymbol{B}, \boldsymbol{R}|\boldsymbol{Q}]\,\mathbb{P}[\boldsymbol{Q}]}{\mathbb{P}[\boldsymbol{B}, \boldsymbol{R}]}$$

Now, $\mathbb{P}[\boldsymbol{B}, \boldsymbol{R}|\boldsymbol{Q}]$ will be 0 unless the quiescent interval on each trial is shorter than the interval between the last two button presses on that trial: a stimulus is only delivered if the subject waits the full quiescent time. $\mathbb{P}[\boldsymbol{B}, \boldsymbol{R}|\boldsymbol{Q}]$ will also be 0 unless the quiescent interval on each trial is longer than the gap between all non-rewarded button presses on that trial (i.e. all except the last two): if not, the reward would then have been delivered earlier than it was. These two constraints define a "feasible range" for the value of $q_t$ on each trial $t$ (Figure 1A). Furthermore, under the null hypothesis, $\mathbb{P}[\boldsymbol{B}, \boldsymbol{R}|\boldsymbol{Q}]$ shows no further dependence on $\boldsymbol{Q}$: the precise value of $q_t$ within the feasible range affects the stimulus time but not reward delivery, and under the null hypothesis the subject cannot detect stimuli. Thus, we may sample from $\mathbb{P}[\boldsymbol{Q}|\boldsymbol{B}, \boldsymbol{R}]$ by resampling $q_t$ on each trial from its original distribution, truncated to that trial's feasible range. For

some choices of prior, this truncation can be computed analytically: for example, for a uniform distribution, the truncated distribution is uniform over the intersection of the original range and the feasible range. If no analytic distribution is available, we can still sample from $\mathbb{P}[\mathbf{Q}|\boldsymbol{B},\boldsymbol{R}]$ by rejection sampling, picking each $q_t$ independently from the original distribution, and regenerating any $q_t$ which do not lie in the feasible range.

To illustrate the method, we constructed simulations where the subject used one of three behavioral strategies (Figure 1B-D). As test statistic we used the mean reaction time: the time between the each trial's stimulus appearance and final button press, averaged over trials.

In the first "random strategy" (Figure 1B-D, top row), the subject presses the button randomly without regard to stimuli or rewards, with inter-press intervals drawn from a Gaussian distribution (800±300 ms, mean ± s.d.). The conditional randomization test rejected the null hypothesis at significance level p<.05 in 48/1000 simulations, as expected when the null is true.

The second "response strategy" (Figure 1B-D, middle row) simulated a subject who could see the stimulus. The subject waits for whichever is shorter: a draw from a fixed Gaussian distribution (800±300 ms, mean ± s.d.), or a draw from a stimulus-dependent Gaussian (stimulus time+150 ±50 ms, mean ± s.d.). The conditional randomization test rejected the null hypothesis in 997/1000 simulations, confirming the test can reject an invalid null.

The third "deceleration strategy" simulated a subject who can detect rewards but not stimuli (Figure 1B-D, bottom row). Following the start of each trial (signaled by reward delivery) the subject pressed the button with progressively longer intervals until the next reward came. The first inter-press interval of each trial followed a Gaussian (150±50 ms, mean ± s.d.) and successive intervals were incremented by adding numbers drawn from this same distribution. This strategy obtains shorter mean reaction times than the second strategy, even though the subject cannot see the stimulus. Nevertheless, the conditional randomization test rejected the null hypothesis only 45/1000 times, the level expected when the null is true.

We conclude that conditional randomization can be safely used for the reaction time task: it rejects the null the simulated strategy depending on stimulus detection, but not for strategies depending only on reward detection.

### Example 2: probabilistic choice task

Our second example is a probabilistic choice task, similar but not identical to that used by the International Brain Laboratory (IBL) [6]. On each trial, a stimulus is presented on either the left or right. The probabilities of the stimuli are unequal and switch in blocks; in a "left block" a left stimulus appears with probability $\alpha$ and a right stimulus appears with probability $1-\alpha$ ($\alpha = 0.8$ for the simulations below), and these probabilities reverse in "right blocks". The blocks switch randomly but the details the block distribution do not affect our current analysis. On each trial the subject chooses left or right, and is rewarded for a choice matching the stimulus side with probability $\beta$, and with probability $\gamma$ for a non-matching choice (for the simulations below $\beta = 0.8$ and $\gamma = 0.2$). Note that in the original IBL task, $\beta = 1$ and $\gamma = 0$; a conditional randomization test cannot be used in this deterministic reward scenario, as discussed below.

A subject who can see the stimulus can obtain a reward on even trial. However, a subject who can detect rewards but not stimuli can still obtain reward on a majority of trials, for example by applying a win-stay lose-switch strategy. To test if the subject can see the stimulus we cannot use a simple randomization test, since this tests whether the entire stimulus sequence is independent of the entire action sequence. Because the subject can detect rewards, they can indirectly infer information about previous stimuli, so the full sequences of stimuli and responses need not be independent even if the subject's choice on trial $t$ does not directly depend on the stimulus on trial $t$.

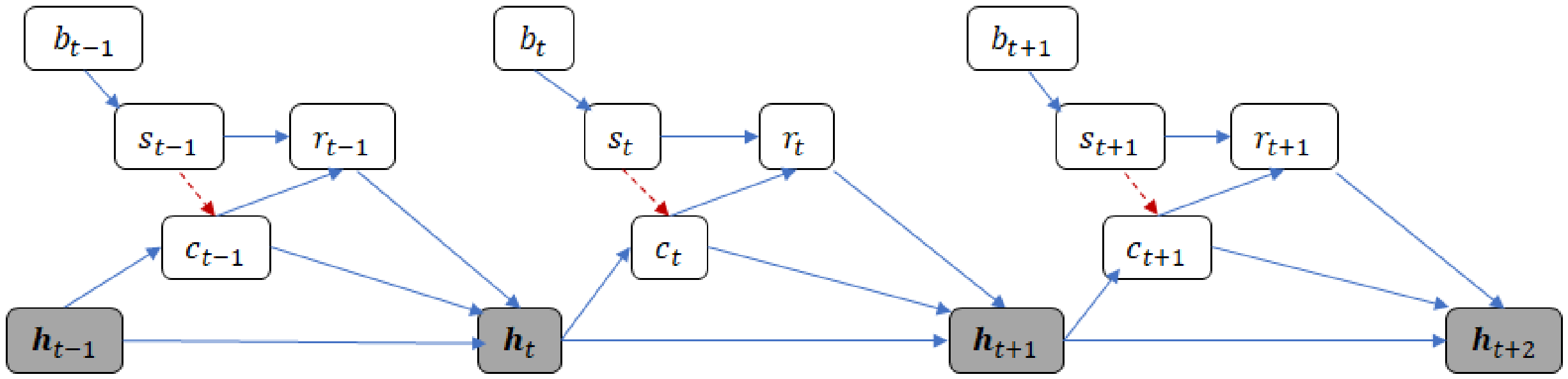

**Figure 2 | Bayesian network for the probabilistic choice task.** The variables on trial $t$ are the block side $b_t$, whose values for all trials are synthesized before the experiment starts; the stimulus $s_t$, which is probabilistically determined by the block; the subject's choice $c_t$, which depends on a "hidden state" $\boldsymbol{h_t}$ representing the subject's high-dimensional and unmeasurable cognitive state (shading indicates non-observability). $\boldsymbol{h_t}$ in turn depends on the cognitive state, choice, and reward on the previous trial. Finally, the reward $r_t$ depends on the choice and stimulus. Blue arrows represent dependence of one variable on another under the null hypothesis that the subject cannot see the stimulus; red dashed arrows represent additional dependence of $c_t$ on $s_t$ under the alternative hypothesis. Under the null hypothesis, $\{b_t, r_t, c_t\}$ form a Markov boundary of $s_t$, meaning that the conditional probability of $s_t$ given all other variables depends only on the values of $b_t, r_t$, and $c_t$.

To design a conditional randomization test for this null hypothesis, we examine a Bayesian network summarizing the probabilistic structure of the task (Figure 2). A Bayesian network is a graphical representation of the dependencies between a set of random variables [7]. A variable in a Bayesian network is independent of all other variables conditional on its *Markov boundary*, composed of its parents, its children, and its children's other parents. We see from Figure 2 that under the null hypothesis, the Markov boundary of $s_t$ is $\{b_t, c_t, r_t\}$, so $s_t$ is independent of all other variables conditional on $\{b_t, c_t, r_t\}$. Thus, the conditional probability of a stimulus sequence $\boldsymbol{S}$ given the sequences of blocks, choices, and rewards factors as

$$\mathbb{P}[\boldsymbol{S}|\boldsymbol{B},\boldsymbol{C},\boldsymbol{R}] = \prod_t \mathbb{P}[s_t|b_t, c_t, r_t]$$

We can thus generate a null ensemble by resampling each $s_t$ independently from $\mathbb{P}[s_t|b_t, c_t, r_t]$. To find $\mathbb{P}[s_t|b_t, c_t, r_t]$, we apply Bayes' rule conditional on $\{b_t, c_t\}$:

$$\mathbb{P}[s_t|b_t, c_t, r_t] = \frac{\mathbb{P}[r_t|b_t, s_t, c_t]\mathbb{P}[s_t|b_t, c_t]}{\mathbb{P}[r_t|b_t, c_t]}$$

Under the null hypothesis, the subject's choice does not depend on the stimulus, so $\mathbb{P}[s_t|b_t, c_t] = \mathbb{P}[s_t|b_t]$ (formally, this follows from the local Markov property of the Bayesian network [7]). Furthermore, because the denominator of the above formula does not depend on $s_t$, we can conditionally resample $s_t$ by normalizing the distribution

$$\mathbb{P}[s_t|b_t, c_t, r_t] \propto \mathbb{P}[s_t|b_t]\mathbb{P}[r_t|b_t, s_t, c_t]$$

To illustrate, consider a trial on which the subject chose left in a left block but was not rewarded. The unnormalized probability that the stimulus was on the left is $\alpha(1-\beta)$, and the unnormalized probability that the stimulus was on the right is $(1-\alpha)(1-\gamma)$. The

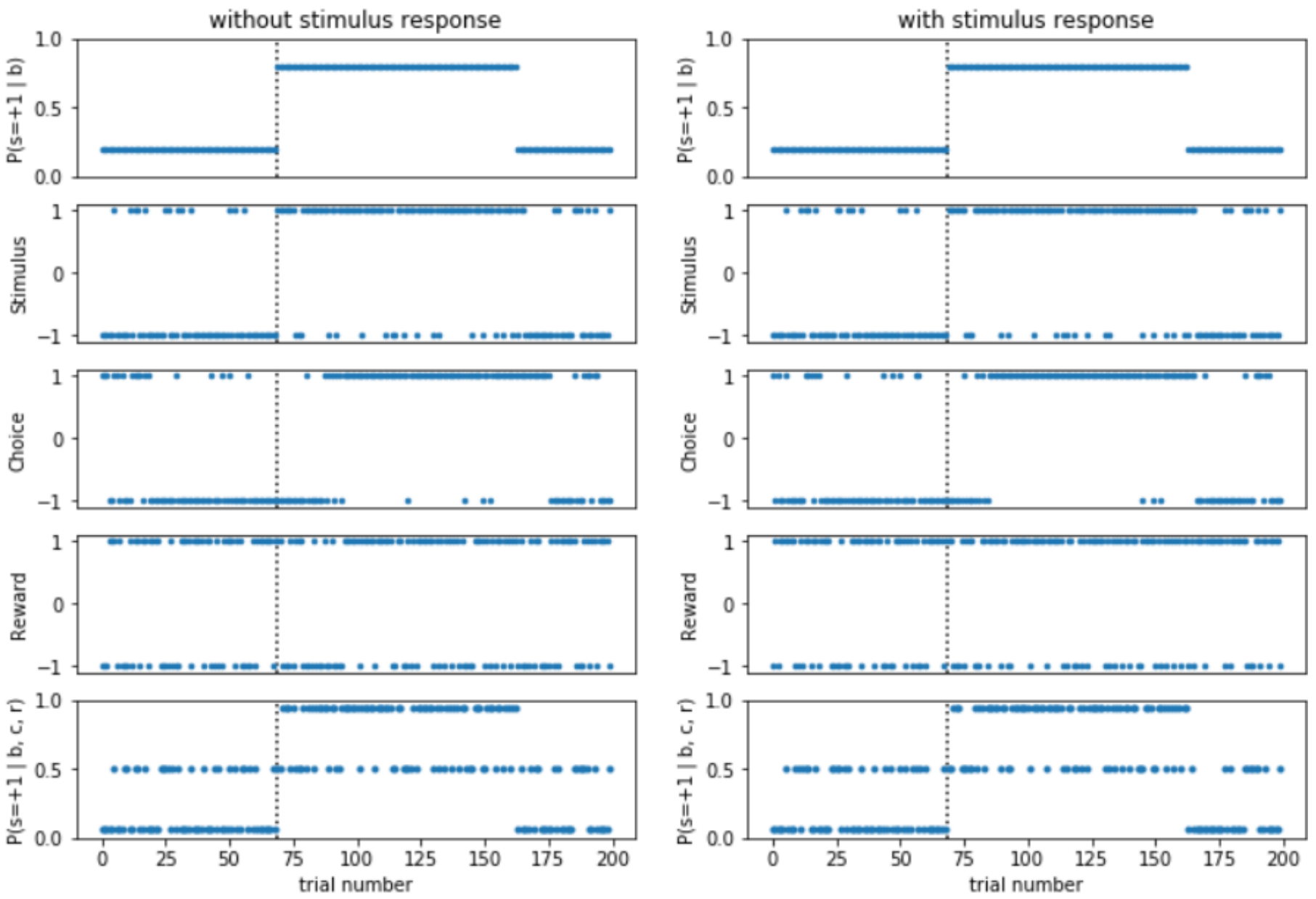

**Figure 3 | Example simulations for the choice task.** Each column shows the first 200 trials from a simulated agent performing the choice task. Left: an agent that cannot see the stimulus, but still learns to make correct choices due to reinforcement learning. Right, an agent that uses both the stimulus and reinforcement learning.

normalized probability of the stimulus appearing on the left is therefore

$$\frac{\alpha(1-\beta)}{\alpha(1-\beta)+(1-\alpha)(1-\gamma)} = \frac{0.16}{0.16+0.16} = 0.5$$

Thus, if the subject chose left in a left block but was not rewarded, the stimulus was equally likely to have been on the left or right. This is because either case requires an unlikely event: either the subject was not rewarded for a correct choice, or the stimulus appeared on the side opposite the block. Both possibilities are equally unlikely, leading to a conditional probability of 0.5 for each stimulus side.

Now consider a trial where the subject was rewarded for choosing left in a left block. The unnormalized probability that the stimulus was on the left is $\alpha\beta$, and the unnormalized probability that the stimulus was on the right is $(1-\alpha)\gamma$. The normalized probability that the stimulus was on the left is thus

$$\frac{\alpha\beta}{\alpha\beta+(1-\alpha)\gamma} = \frac{0.64}{0.64+0.04} \approx 0.94$$

If the subject was rewarded for choosing left in a left block, it is thus very likely that the stimulus was on the left; the only way the stimulus could have appeared on the right is if two unlikely things happened: a right stimulus appeared in a left block, and then a reward was given for an incorrect choice.

We simulated a subject's performance of this task using "mixture of agents" model [8] (Figure 3, left). The model combines reinforcement learning term, which makes the subject more likely to repeat a previously-rewarded choice, and a "habit" term, which causes the subject to repeat previous choices even if not rewarded Although the simulated subject could not directly detect the stimulus, they still performed the task reasonably accurately, switching choices fairly soon after block transitions. A second simulated agent's choices were based on this learning model combined with direct stimulus detection (Figure 3, right). Full details and parameters can be found in the online code.

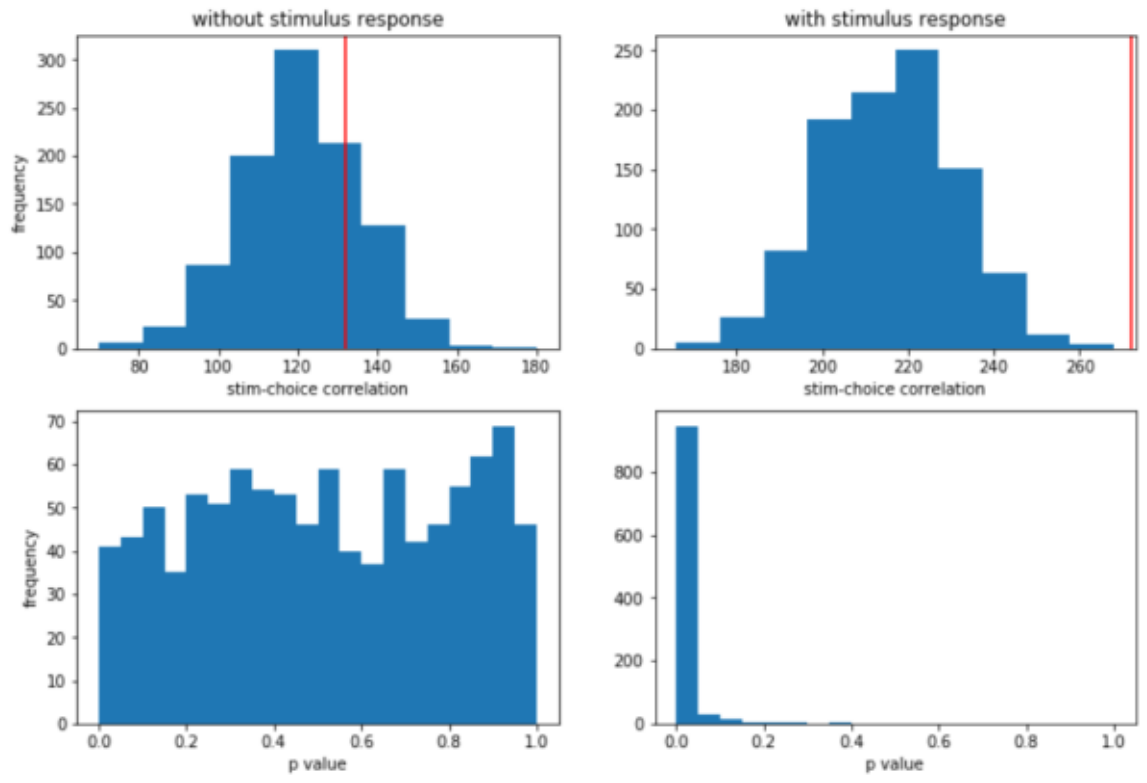

**Figure 4 | Conditional randomization test for the choice task.** Top row: null ensemble of test statistic (blue histogram) and actual data statistic (red line), for example sessions of the two agents. Bottom row: histogram of p-values for 1000 simulated sessions of the two agents.

The conditional randomization test correctly rejected the null hypothesis of no stimulus detection for the second, but not the first agent (Figure 4). To show this, we used as test statistic the correlation between stimulus and choice on each trial ($\sum_t c_t s_t$). The null hypothesis was rejected in 41/1000 simulations of agent 1, and 948/1000 simulations of agent 2.

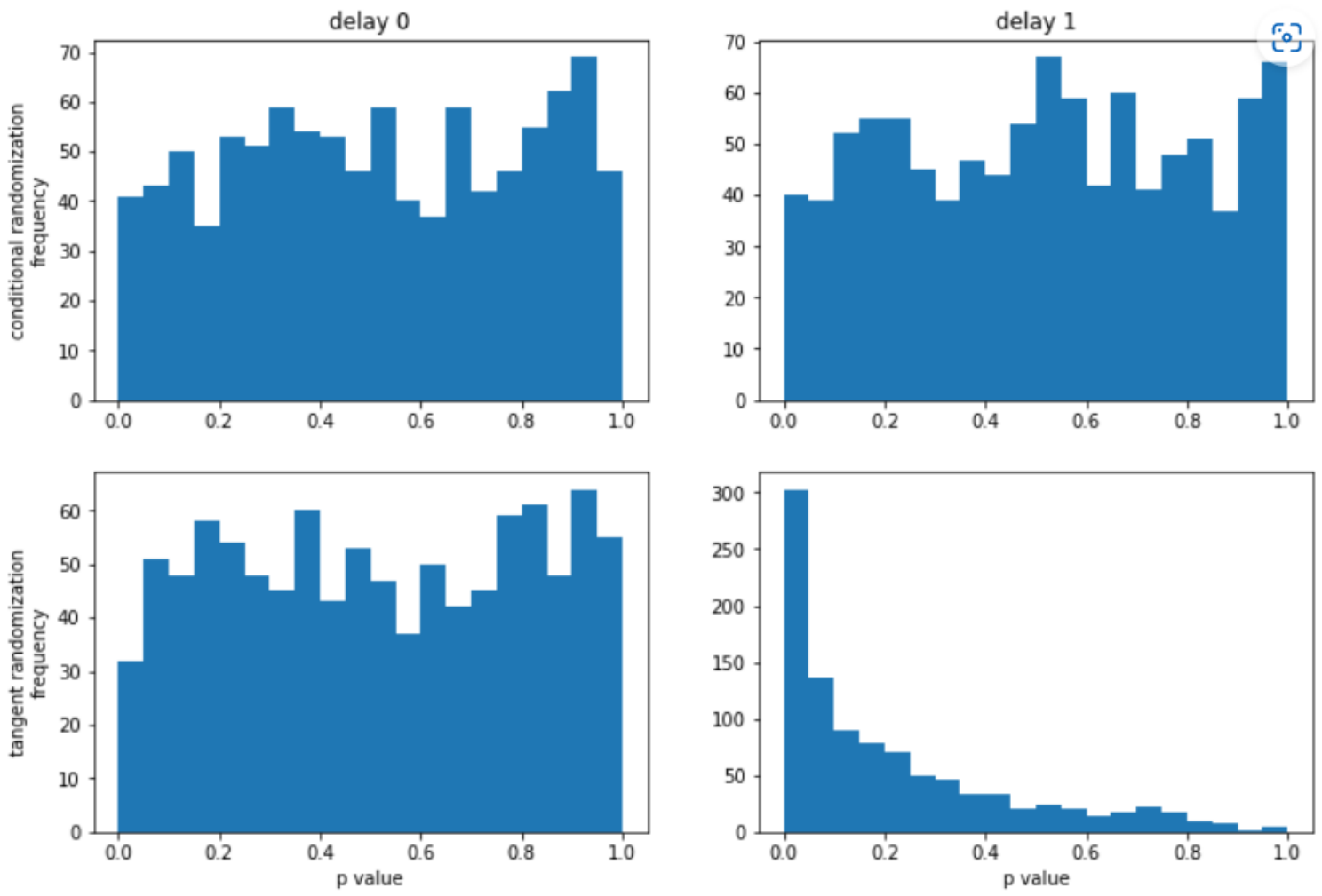

**Figure 5 | Conditional randomization vs. tangent randomization.** Each panel shows a histogram of p-values for 1000 simulations of the choice task, for a simulated agent which cannot see the stimulus, so the null is valid. The left column uses test statistic $\sum_t s_t c_t$, and the right column uses $\sum_t s_t c_{t+1}$. The top row uses conditional randomization, the bottom row uses tangent randomization. Conditional randomization gives correct results for both test statistics; tangent randomization however falsely rejects the null when using the second test statistic.

## Conditional randomization vs. tangent randomization

In the conditional randomization tests we have just described, we resample randomized experimental variables (quiescence interval, stimulus side) conditional on other events (rewards, choices) that occur in both the past and future. This is necessary because the future behavior of the experimental apparatus depends on the randomized variables even if the subject is unable to detect them. Conditioning "backwards in time" is necessary to ensure that the resampled variables are consistent with the future behavior of the experimental apparatus.

An alternative approach that one might consider is to resample the experimental variables conditional on only events that happened earlier in time. A sequence generated in this way is called a *tangent sequence*[9]. One can construct a statistical test by *tangent randomization*, where the actual value of a test statistic is compared to a null ensemble generated from hypothetical experiments where randomized experimental variables are replaced by resampled tangent sequences.

Tangent randomization sometimes, but not always, gives a valid test[10]. A true randomization test requires that the probability of observing the randomized data equals the probability of observing the original data; if this holds then the test is exact for any test statistic, and for any dataset size. Conditional randomization tests satisfy this requirement, but tangent randomization tests do not. In fact, the probability of actually observing a resampled tangent sequence is often 0, for example if a stimulus was resampled that is consistent with the past but inconsistent with the future state of the experiment. Nevertheless, with some choices of test statistic and in the limit of long time series, tangent randomization becomes equivalent to the *Martingale Z-test* [10], which is asymptotically conservative: as the length of the timeseries tends to infinity, the probability of rejecting a valid null hypothesis will not exceed the significance level. The requirements for this equivalence to hold are described in Ref. [10] and will not be repeated here. Instead we will give two examples, with conditional randomization working on both, but tangent randomization on only one.

Both examples use the choice task of the previous section. Tangent randomization in this case is simple: to randomize $s_t$ conditional on preceding history one simply draws it from $\mathbb{P}[s_t|b_t]$, giving an 80% chance of a left stimulus in a left block, and so on. We simulate data for which the null hypothesis that the subject cannot see the stimulus is valid, using the same mixture of agents model as above.

For our first example the test statistic is the same as before: $\sum_t s_t c_t$. Tangent sequence randomization is valid with this test statistic: the null is rejected in 32/1000 simulations, compared to 41/1000 for the conditional randomization test (Figure 5, left).

For the second example the test statistic is the correlation of the stimulus with the choice one trial delayed: $\sum_t s_t c_{t+1}$. Even though the subject cannot see the stimulus, there is a causal effect of $s_t$ on $c_{t+1}$, since $s_t$ affects the reward $r_t$, which the agent perceives prior to making the choice $c_{t+1}$. The conditional randomization test is still valid with this test statistic, rejecting the null 40/1000 times. The reason for this is that $s_t$ is resampled conditional on $r_t$, so the resampled stimuli are consistent with the reward that is later delivered. However tangent randomization fails in this case, rejecting a valid null 303/1000 times (Figure 5, right). We conclude that while tangent randomization is sometimes valid, it's validity depends on the test statistic and must be checked on a case-by-case basis. In contrast, conditional randomization is valid for any test statistic.

## Conclusion

Conditional randomization can often be used to produce exact tests in experiments where some variables are dynamically randomized depending on a subject's previous behavior. Conditional randomization can be used whenever it is possible to derive a conditional distribution for some of the randomized experimental variables given other observations. Deriving these distributions can sometimes be difficult, but if the conditional distribution is derived correctly then the resulting test is very safe to use: like an ordinary randomization test it is exact, guaranteed to reject the null at the specified rate, for any choice of test statistic and any sample size, not requiring any assumptions except knowledge of the rules used to randomize the experiment. This is not true of tangent randomization, which works only for appropriate choices of test statistic and sufficiently large sample sizes[10].

Conditional randomization cannot be used in all circumstances, and whether it can be used for a given experiment must be determined on a case-by-case basis. Example 1 presented a simplified version of test first used in Ref. [5]. We had not designed the experiment planning to use a conditional randomization test. Instead, we derived the conditional randomization test for this task only after failing to find any other way to test the null hypothesis; the fact that this was possible was fortuitous. In some cases, conditional randomization cannot be used at all. For example, the IBL task[6] is the same this paper's Example 2, with deterministic rather than probabilistic rewards ($\beta = 1$ and $\gamma = 0$). However the conditional randomization test we derived could not be used in this case, as the resampled stimulus sequences would be same as the originals.

Conditional randomization is a useful addition to the arsenal of an analyst of neural and behavioral data. In an ideal world, experiments would be planned with data analyses in mind, and experiments could be designed a priori to allow methods such as conditional randomization. In the real world this is not always possible. However, if experiments are designed with as many variables as possible randomized, it increases the chance that methods like conditional randomization can be used.

### Code availability

Python code for examples 1 and 2 can be found at https://zenodo.org/records/10061296, and can be run online at https://colab.research.google.com/drive/1QbJx0fxvsk_JWc72K19QAlAtP_XItERJ and https://colab.research.google.com/drive/1tQH9Ws0pgO30k363Zt6ldvAU2fCxIKvA

### Acknowledgements

This work was supported by the Wellcome Trust (223144/Z/21/Z) and ERC (694401).